# Artificial Dust Based Attack Modelling: A Threat to the Security of Next Generation WCN

Misbah Shafi, *Student Member, IEEE*, Rakesh Kumar Jha, *Senior Member, IEEE*

*Abstract*— This paper introduces a systematic and novel mechanism for devising a security attack in the WCN (Wireless Communication Network). The proposed model involves the implementation of the AD (Artificial Dust) by the intruder, followed by the execution of the HD (Half-Duplex) attack. The communication network is based on the deployment of urban and rural scenarios with an unknown CSI (Channel State Information). Depending on the achieved path loss based on the distance of the user from the BS, the user with the highest path loss is particularized for the attack. The formulation of AD divulges the increased susceptibilities of the secure network specifically for the selected legitimate user. The parameter of visibility defines the amount of AD present in the communication channel. Based on the enumerated attenuation created by the artificial dust, the parameter of secrecy rate is evaluated with varying distance of the user from the BS and the operating frequency. Furthermore, the proposed scheme of the HD attack is initiated by the intruder at the specified valid user. The strategy of the attack focuses on the continuous monitor of the uplink and attempts the spoofing attack on the downlink wherein the allocation of the resources takes place. The efficacy of the proposed approach is corroborated through the examination of simulation results. The assessment of the proposed mechanism highlights notable characteristics as compared to the conventional methodology of the FD (Full-Duplex) attack.

*Index Terms*— AD (Artificial Dust), HD (Half-Duplex) attack, secrecy rate, security, attenuation, path loss, WCN (Wireless Communication Network)

## I. INTRODUCTION

Next generation Wireless Communication Network (WCN) is predominantly progressing and evolving a wide range of developments in terms of applications, services, and efficacy. It has equipped a significant contribution to the life of the people for the transmission of confidential information such as health care data, credit card records, energy pricing, control, and command information [1]. For the next-generation WCNs, the increasing requirement of broadband services has attained the spectrum exhaustion for carriers. This challenge is addressed by the utilization of the underused spectrum defined by the mmWaves (Millimeter-Waves) ranging from 30GHz to 300GHz. The next generation WCN incorporates B5G / 6G networks. However, incorporating millimeter Waves occupies various limitations. One of the most significant obstacles is the path loss attained by the mmW signals [2]. The increased path loss is due to the increased impact of attenuation.

### A. Related work

Propagation of the microwaves in the presence of dust creates an attenuation impact, in the form of absorption loss and scattering loss of the transmitted signal [3]. Various calculations based on the dust attenuation interfering in the communication are defined by Mie scattering and Rayleigh scattering. However, Rayleigh scattering is defined for the particles smaller than the wavelength, and Mie scattering is defined for the effects of scattering and absorption for spherical particles [4]. In [5], it was described that the cross-polarization predominantly disturbs wave propagation around the frequency of 10GHz. The utmost attention of the previous work is mainly concerned with cross-polarization, scattering, and attenuation. The analysis of the parabolic wave equation was defined to execute the effects of the scattering and absorption by the presence of the dust particles in dust and sand storms. [6]. For the earth satellite communication link, the presence of dust and sand storms has a direct impact on microwave propagation. The existence of the phase shift and cross-polar discrimination due to the presence of the dust particles has a direct impact on height, elevation angle, visibility, and frequency. Consequently, affects the propagating signal [7]. For the analysis of a three-dimensional scenario of dust and sand storms parabolic wave equation methodology is utilized with the inclusion of ground reflection and the effects of earth's curvature on the propagating microwave propagation of the transmitting signal. The analysis is considered under the study of the height of the transmitting antenna and visibility profile [8]. For the radar communication scenario, parabolic wave equation analysis is defined to estimate the attenuation in the form of backscattering power. The inspection of the backscattered power is evaluated based on the approximation of the Mie scattering under the considerations of beam divergence and the earth's curvature [9].

Most of the prevailing work on dust in the microwave and millimeter-wave propagation considers the signal strength of the transmitted signal as the parameter of execution. The impact of dust attenuation in the field of security is identified as the new research direction and occupies primary importance, especially in the communication networks of defense and healthcare. It provokes prominence for the determination of the security impact on the next generation wireless communication

Misbah Shafi is with the department of electronics and communication, Shri Mata Vaishno Devi University, Katra, India 182230, Rakesh Kumar Jha is with the department of electronics and communication, Indian Institute of

Information Technology Design and Manufacturing Jabalpur, India 482005. (email: misbahshafi0@gmailcom, jharakesh.45@gmail.com)

TABLE I
COMPARISON OF THE PROPOSED SCHEME WITH THE EXISTING ATTACKING METHODOLOGIES

| Ref. | Objective | Associate parameters | Type of an attack | Supported application | Security analysis of AD | Attack target (UL/DL) | Motivation |
|---|---|---|---|---|---|---|---|
| [11] | The attacker poisons the data of spectrum sensing at the transmitter end by communicating during the idle time slots of the transmitter. | Mis-detection, false alarm, throughput | Spectrum poisoning attack | Spectrum sensing | ✗ | UL,DL | **Inadequacies of [11]-[14]:** Application specific, not applicable to next generation wireless communication network, high miss rate, less sensitivity, more vulnerable to attack detection |
| [12] | An intelligently reflecting surface is used to reflect signals from the valid transmitter to the valid receiver such that the received signals from reflecting links and direct links are added destructively, thereby deteriorating the SINR (Signal to Interference Noise Ratio). | SINR (Signal to Interference Noise Ratio) | IRS (Intelligent Reflecting Surface) jamming attack | Wireless communication network, IoT (Internet of Things), beamforming | ✗ | UL,DL | |
| [13] | The attack operates the coupling vulnerability in presence of the hidden nodes. The execution of the attack is propagated beyond the start of the location for longer time periods to force the network to be operated at the lowest bit rate. The resultant change creates a transition from an uncongested state to a congested state. | Throughput, node utilization | Cascading attack | WiFi-based communication network, HetNet, linear network | ✗ | UL,DL | **Adequacies of the proposed scheme:** Applicable to beyond 5G, less miss rate, more sensitivity, less vulnerable to attack detection, a threat to the communication network of defense, future networks such as augmented networks, telesurgery |
| [14] | The comparable signals are transmitted to the keying devices such that the attacker manipulates the channel measurements to compromise the portion of the key. | Attacker energy factor, detection rate, false alarm rate, key generation rate, normalized signal strength | Manipulative attack | Wireless communication network | ✗ | UL,DL | |
| Proposed | The security impact of AD (Artificial Dust) followed by the execution of HD (Half-Duplex) attack | Miss-rate, Dust attenuation, secrecy rate, sensitivity, energy efficiency, complexity | Half Duplex attack | D2D, UDN, VANETs, IoT, Spectrum sharing | ✓ | DL | |

network in the presence of AD (Artificial Dust) [10]. The concept of AD is utilized instead of real dust because the AD is executed based on the interest of the intruder and consequently the concentration of AD is increased or decreased.

Various attack models were developed for conventional Wireless Communication Network (WCN) based on the Channel State Information (CSI). However, with the development of 5G / Beyond 5G / 6G WCN the channel conditions and the other associated parameters such as spectrum, frequency, power have been changed. Therefore, the possible attack scenarios applicable to the emerging technologies of WCN are required to be addressed. The recent attacks [11]-[14] evaluates the parameters such as throughput, SINR (Signal To Interference Noise Ratio), and normalized signal strength. Moreover, these attacks involve both uplink and downlink for the attack to take place. Considering all these parameters a comparative analysis is made between the recent attacks [11]-[14] and the proposed attack model as shown in Table I. Artificial Dust-based attack is identified as the latest possible research direction in the upcoming WCN. The attack can prove catastrophic in the applications of next generation WCNs especially in the fields of defense and healthcare. The proposed attack modeling involves the technology of D2D communication. In addition to that, the attack can also be executed in the technologies of spectrum sharing, VANETs (Vehicular Ad-hoc Networks), IoT (Internet of Things), and UDN (Ultra Dense Network). An assessment of the current schemes incorporating different attacking methodologies in the communication network and the proposed methodology of the security aspect of AD (Artificial Dust) executing HD (Half-Duplex) attack is summarized in Table I.

### B. Contribution

This paper defines the attack modeling in the security framework of next generation wireless communication networks. The attack is initiated by the background of the AD followed by the execution of the Half-Duplex (HD) attack to attain resource spoofing via this work. The fundamental contributions of this paper are summarized as follows:

1) We have proposed an attack modeling based on the impact of Artificial Dust. The user with maximum path loss is targeted for the security attack by the intruder. The execution of the AD is made by the intruder around the targeted user

2) The HD attack is performed on the selected user, such that resources meant to be allocated to the selected user are spoofed by the intruder

*3)* The proposed methodology of HD attack is compared with the FD (Full-Duplex) attack in terms of throughput analysis, secrecy rate, sensitivity, probability of missed attempts, and complexity. to determine the effectiveness of the attack

The organization of the paper is as follows. Section II illustrates the system model. Section III demonstrates the mechanism of the HD attack. The representation of the proposed scheme is interpreted in Section IV. Section V depicts the attacking model through an illustrative example. The performance of the mechanism is evaluated through the simulation results given in Section VI. Finally, the conclusion of the paper is stated in Section VII.

## II. SYSTEM MODEL

The description of the system model incorporating Artificial Dust (AD) followed by the execution of the HD (Half-Duplex) attack is presented in this section. Further, the identification of the problem formulation is stated in this section.

### A. Artificial Dust

Artificial dust is defined as the phenomenon of creating dust in the atmosphere in the form of suspended small dust particles. The process can be operated by the ground base generator or via rocket or through an aircraft. The artificially generated dust particles can also be dispersed in the atmosphere by UAV or remote-controlled rockets depending on the feasibility of the operation. The artificial dust particles include solid carbon dioxide, sulphates, ammonium containing aerosols, nitrates, and fly ash particles. Based on particle size the artificial dust particles can be categorized into two types viz. primary sized particles and secondary sized particles. The primary sized particles are of the size range of $5\text{-}35\mu m$. The size of secondary-sized artificially generated dust particles lies in the range of $2\text{-}5\ \mu m$. CARE (Charged Aerosol Release Experiment) is one of the experimental AD projects operated by NASA to witness the characteristics of artificial dust clouds known as Noctilucent Clouds.

### B. System model

The framework of the communication scenario in the presence of the AD for the attack modeling of the HD attack is portrayed in Fig. 1. The schematic of the scenario consists of $i$ number of valid users, out of which $k$ D2D (Device to Device) pairs are served by the BS (Base Station) present at a coverage radius $r = 250m$. These D2D pairs are labeled as $D_1, D_2, D_3, ..., D_k$ such that $2k < i$. Consider an eavesdropper is present at the edge of the cell and is capable of monitoring the communication network. In this proposed scheme, the communication network devoid of the eavesdropper without the effect of AD is compared with the communication network under the impact of eavesdropper and AD attenuation. Assume $h^k$ as the channel coefficient from BS to the $K^{th}$ D2D pair having 2 users labeled device-1 and device-2 and $h_{AD}^k$ as the channel coefficient from BS to the same D2D pair under the impact of AD in the presence of an eavesdropper such that $h^k > h_{AD}^k$.

**Remarks 1:** *The channel state information (CSI) is not known to the intruder. However, the intruder is familiar with the location of the valid user, from which the resultant path loss is computed. Considering all the users are moving with a negligible speed and are examined as static for the respective timestamp of execution.*

Based on the achieved path loss, the D2D pair with the most significant path loss is targeted for the attack. The transmission power from the BS be $p_{Bs}$, such that the received power $p_{D_{k1}}$ by the device-1 in the $K^{th}$ D2D pair is defined as :

$$p_{D_{k1}} = p_{Bs} w \left[\frac{l_o}{l_b^{k1}}\right]^\varsigma \tag{1}$$

where $w$ is a dimensionless constant. It depends on the characteristics of the average attenuation of the channel. It can be expressed as :

$$(w)_{dB} = 20 log_{10} \frac{\lambda}{4\pi l_o} \tag{2}$$

$l_o$ is the reference distance, $l_b^{k1}$ is the distance from BS to the device-1, and $\varsigma$ is the path loss exponent, $\lambda$ depicts the wavelength of the radiation. From equation (1) the free space path loss at device-1 $(p_L^{k1})_{dB}$ can be evaluated as :

$$(p_L^{k1})_{dB} = 10\varsigma log_{10} \left[\frac{l_b^{k1}}{l_o}\right] - 10 log_{10}(w) \tag{3}$$

Incorporating the effect of shadow fading in the free space path loss in equation (3) [21] we get :

$$(p_L^{k1})_{dB} = 10\varsigma log_{10} \left[\frac{l_b^{k1}}{l_o}\right] - 10 log_{10}(w) + \gamma_{dB} \tag{4}$$

Similarly, associating the impact of the AD such that the attenuation of AD incorporated in path loss $(p_L^{k1})_{dB}^{AD}$ at the device-1 of the $k^{th}$ D2D pair is expressed as :

$$(p_L^{k1})_{dB}^{AD} = 10\varsigma log_{10} \left[\frac{l_b^{k1}}{l_o}\right] - 10 log_{10}(w) + \gamma_{dB} + \beta_{k1}^{AD} \tag{5}$$

where $\beta_{k1}^{AD}$ is the attenuation due to the AD at the $k^{th}$ D2D pair for the device-1. The strength of the propagating signal in the presence of dust particles is drastically affected due to absorption and scattering losses. These losses are interdependent on the factors of dielectric constant, frequency of operation, visibility, and the radius of the dust particles. Based on the direct and inverse relationship between the AD attenuation and these factors, the proposed model is attained by combining these factors with the optimization of the constants. The proposed model for the calculation of the AD attenuation on the propagation of the signal can be expressed as:

$$\beta_{k1}^{AD} = \frac{x r_e \varepsilon^{"} f}{V\left[(\varepsilon'+2)+\varepsilon^{-2}\right]} [dB/Km] \tag{6}$$

where $x$ is a constant. It can be obtained and optimized from the calculations of the measured attenuation. The measured attenuation can be attained from the equal approximation of the received signal with the transmitted signal concerning the path loss. From equation (6), $x$ is obtained as:

$$x = \frac{\beta_{k1\ Measured}^{AD} V\left[(\varepsilon'+2)+\varepsilon^{-2}\right]}{r_e \varepsilon^{"} f} \tag{7}$$

The SNR in the presence of the AD $(\varpi)_{AD}$ is given by:

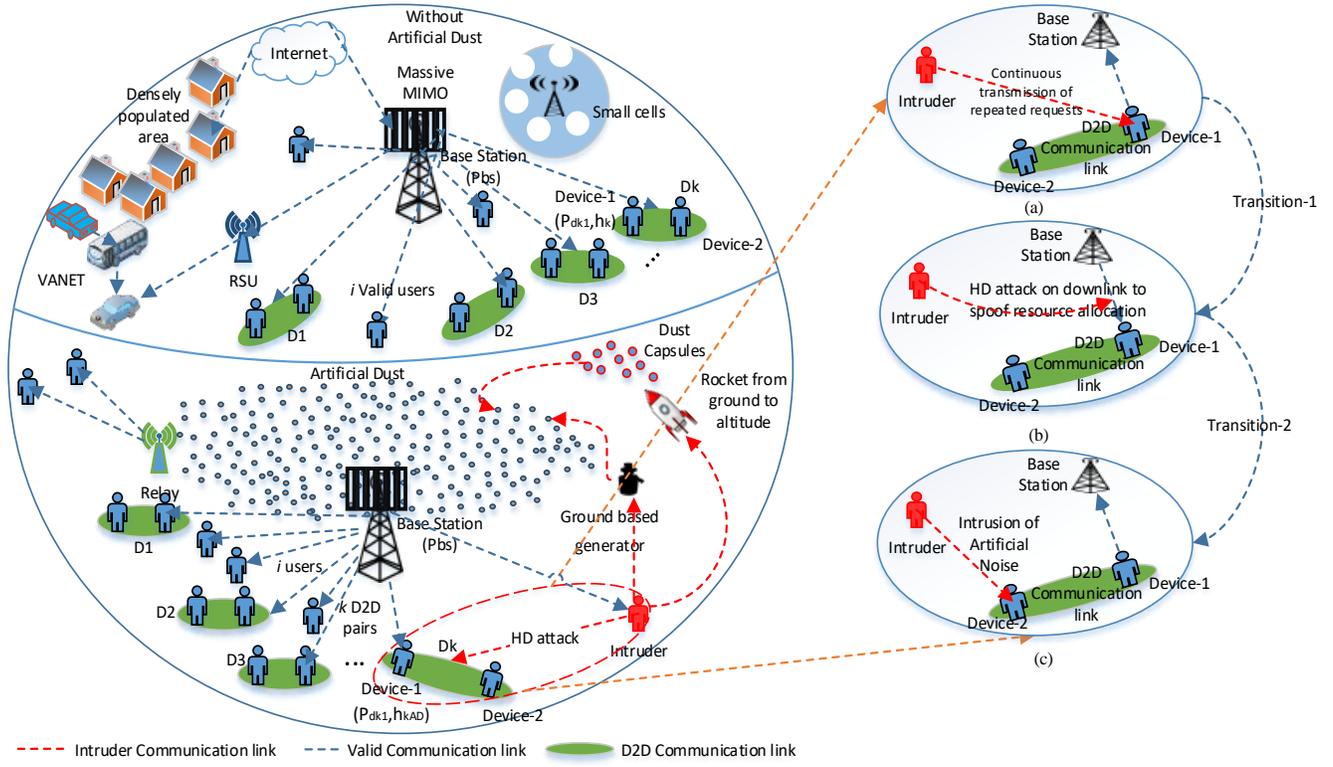

Fig. 1. System model.

$$(\varpi)_{AD} = \frac{h_{AD}^{k1} P_{Bs}}{\kappa} \qquad (8)$$

where $h_{AD}^{k1} = 10^{-\frac{(P_L^{k1})_{dB}^{AD}}{10}}$, $\kappa$ is the noise power. The required capacity of the device-1 for the $k^{th}$ D2D pair is obtained as:

$$(C_{k1})_{AD} = B\log_2(1 + (\varpi)_{AD}) \qquad (9)$$

Secrecy rate estimates the secure transmission defined by the capacity which is dependent on the SNR. Considering secrecy rate as the evaluating parameter of security, the secrecy rate under the impact of AD defines the effectiveness of the attack. Therefore the secrecy capacity can be represented as:

$$(C_S)_{k1} = [(C_{k1})_{AD} - (C_{ev})]^+ \qquad (10)$$

Analyzing the above equations (5), (8), and (9) the influence of AD increases the path loss. It creates a drastic decrease in the SNR. Consequently, results in a deterioration of the capacity of the valid user. Therefore, from equation (10), it is quite evident that the presence of AD has a direct impact on the security of the communication network.

**Proposition 1:** *The capacity of the device-1 for the $k^{th}$ D2D pair is reduced below the capacity of the intruder due to the impact of the AD on the propagating signal.*

$$(C_{k1})_{AD} < (C_{ev}) \qquad (11)$$

**Proof:** The corresponding capacities of the device-1 and the eavesdropper is given by:

$$(C_{k1})_{AD} = B\log_2(1 + (\varpi)_{AD}) \qquad (12)$$

$$C_{ev} = B\log_2(1 + (\varpi)_{ev}) \qquad (13)$$

where $(\varpi)_{ev}$ is the SNR of the eavesdropper. The relation between the channel characteristics of the eavesdropper and the valid user can be expressed as:

$$(h_{AD}^{k1}) < (h_{ev}) \qquad (14)$$

In the presence of the AD, the SNR for the eavesdropper and the device-1 is illustrated by using equation (14) as:

$$\frac{h_{AD}^{k1} P_{Bs}}{\kappa} < \frac{h_{ev} P_{Bs}}{\kappa} \qquad (15)$$

$$(\varpi_{k1})_{AD} < (\varpi)_{ev} \qquad (16)$$

$$(C_{k1})_{AD} < (C_{ev}) \qquad (17)$$

According to Wyner's wiretap model [19], the secrecy of the network is maintained only, when the capacity of the valid user is greater than the capacity of the eavesdropper. The capacity defines the signal strength of the respective user. Therefore, the intruder is now capable of attacking the communication network as the capacity of the eavesdropper is greater than the capacity of the valid user in the presence of the AD. ∎

## III. GENERALIZED MECHANISM OF HALF-DUPLEX ATTACK

### A. Overview of the proposed attack

The suggested approach of attack modeling includes D2D communication employing the configuration of device relaying controlled link, supported by the BS. Considering the vulnerability of the relay device the attack is initiated on it. The proposed methodology introduces the attack on downlink exclusively to spoof the resources is contrary to the FD attacks wherein both uplink and downlink are aimed for the attack. The dominant improvement of the proposed mechanism is the attack devoid of the authentication process. The authentication process is defined as the security mode commands during the configuration of User Equipment (UE) from Radio Resource

| $x$ | $f(GHz)$ | $\varepsilon'$ | $\varepsilon''$ | $V(Km)$ | $\beta_{k1\ Measured}^{AD}$ |
|---|---|---|---|---|---|
| $2.993 \times 10^{-4}$ | 7.5 | 4.68 | 0.38 | 0.15 | 0.025 |
| $8.58 \times 10^{-4}$ | 13 | 3.9 | 0.63 | 0.05 | 0.67 |
| $4.76 \times 10^{-4}$ | 40 | 4 | 1.3 | 1.4 | 0.069 |
| $2.99 \times 10^{-4}$ | 100 | 3.5 | 1.64 | 0.001 | 180 |

TABLE III
VARIABLE DEFINITIONS

| Symbol | Description |
|---|---|
| $r_e$ | Radius of the dust particle in micrometers |
| $\varepsilon', \varepsilon''$ | Real part of the dielectric constant, imaginary part of the dielectric constant of the dust particle |
| $f$ | Operating frequency in GHz |
| $V$ | Visibility in Km |
| $\beta_{k1}^{AD}, \beta_{k1\ Measured}^{AD}$ | Calculated attenuation of the AD, Measured attenuation of the dust in dB/Km |
| $h_{AD}^k, h_{ev}$ | Channel gain from BS to the device-1 for the $k^{th}$ D2D pair in the presence of AD, channel gain of eavesdropper from BS |
| $(C_{k1})_{AD}, (C_S)_{k1}, C_{ev}$ | Capacity, secrecy capacity of the device-1 for the $k^{th}$ D2D pair in the presence of AD, the capacity of the eavesdropper |
| $\phi_{sc}, \phi_{abs}$ | Scattering efficiency, absorption efficiency |

Control (RRC) idle state to UE-RRC connected state. These commands are exchanged between the UE and the respective gNB (g-Node B). Further, the reduction in the probability of miss-rate can be accomplished by the HD attacking phenomenon. The framework of the proposed methodology is shown in Fig. 1. The CSI of the valid users is not known to the eavesdropper. The device-1 of the $kth$ D2D pair is present at a distance where the strength of the signal from the BS is reachable. The device-2 of the same D2D pair is present at a distance where the signal strength from the BS is considerably weak. The intruder monitors the communication process of the $kth$ D2D pair. After the completion of the authentication process employed by device-1, the intruder targets the downlink, where the resource allocation is performed. The process of the AN (Artificial Noise) is operated by the intruder for the prevention of further requests from the selected D2D pair.

### B. Process of the proposed attack

The approached methodology incorporates the AD attenuation to aim the half-duplex attack. Under this aspect, the capacity of the intruder is greater than the capacity obtained by the valid user in the presence of the AD. The attack is performed on the downlink instead of the uplink. The uplink process conducts the authentication. It reduces the chance of detection of the attack. The resources meant to be allocated for device-1 are spoofed by the intruder while attacking at the downlink. There are three main objectives of the proposed attack. The first objective is to use the identity of the valid user. The second objective is the spoofing of the resources. The third objective is to reduce the chance of intrusion detection. The proposed procedure provides the advantage of minimizing the miss rate probability $P(m_r)$ and to maximize the effectiveness $A_p$ of the attack shown as:

$$A_p \propto \frac{1}{\min\{P(m_r)\}} \qquad (18)$$

The effectiveness of the attack is determined by the parameters of capacity, secrecy rate, sensitivity, miss rate probability, and complexity. Capacity determines the signal strength of the valid user or an intruder. Therefore, decides the favorable attacking condition for the attacker as given in proposition 1. The secrecy rate estimates the secure communication of the network. Consequently, it is considered as a deciding parameter for the attempt of an attack to occur. Sensitivity defines the minimum optimal value of capacity required by the intruder to trigger the attack. Thus, more the sensitivity, minimal is the value of capacity required by the eavesdropper, and thereby more efficient is the attack. The miss rate is defined as the number of missed attempts of an attacker to attempt the spoofing of resources from the valid user and the corresponding probability is termed as a miss-rate probability. Hence, the more effective the attack less is the miss rate probability. The complexity of an attack defines the computations of the attack. The computations determine the usage of a resource for the attack to take place. Thus, for an effective attack, less complexity of an attack is required. The attack can be launched by evaluating the following phases as shown in Fig 1(a), 1(b), 1(c)

#### 1) The phase of repeated request transmission

The main objective of this phase is to make the relay device-1 unapproachable to the downlink from the BS and use the identity of the relay device. It is considered that the eavesdropper is capable of monitoring the communication between the $kth$ D2D pair and the BS. The authentication of the valid user device-1 is completed during uplink. During the phase of downlink, the intruder continuously transmits the requests such that for each request, a response is transmitted by the device-1. The process continues with a practice of requests by the intruder and responses by device-1 of the $kth$ D2D pair. It leads to the overload of the network connectivity. Therefore, drain of resources at the valid user end. Further, the identity of device-1 is spoofed and is impersonated by the respective intruder. The design flow of the request and response can be expressed as follows:

$$R_k = \sum_{i=2}^{k}(h_{ev1}P_i' + A_k) \qquad (19)$$

The received response at the intruder is given by:

$$r_k = \sum_{i=2}^{k}(h_{1ev}p_i + A_k) \qquad (20)$$

The intruder launches the request to device-1 in the form of a random sequence number with the prescribed bytes occupied to prevent the buffer flow due to the extended byte. The requests are transmitted at the same instant to prevent the availability of the device from detection. $R_k$ and $r_k$ are the request and the achieved response by the intruder in the form of a packet with the size defined by the internet protocol. Usually, the size of a request packet is 54 bytes. However, the request transmitted to the target device-1 is the artificial noise (random sequence). The response packets contain the identification number and the sequence number. Where $R_k$ is the summation of $k$ received requests by device-1 for the availability. $A_k$ is the AWGN (Additive White Gaussian Noise) with zero mean and $v^2$ variance $A_k \sim \mathbb{N}(0, v^2)$, $P_i'$ is the transmitted request by the intruder, $h_{ev1}$, $h_{1ev}$ depicts the channel coefficients from eavesdropper to the device-1 and vice versa, $r_k$ is the summation of $k$ received responses by the intruder, $p_i$ is the transmitted response by the device-1.

### 2) The phase of resource spoofing

The fundamental objective of this phase is to spoof the resources of the valid user effectively by considering the reduced probability of miss rate. The reduced miss rate is achieved by targeting the downlink only. After the completion of repeated request transmissions, device-1 is made unavailable for the reception of the downlink from the BS. Instead of device-1, a downlink from the BS is received by the intruder with the identity of the valid user device-1. The successful reception of the downlink defines the fact that the resources are allocated to the intruder with the identity of device-1. Therefore, the intruder employs the attack on the downlink to spoof the resources devoid of the authentication process. The attack solely on the downlink reduces the probability of the miss rate as the number of attempts are aimed at the downlink only. In the case of the FD attack, the same number of attempts are distributed for uplink and downlink, which increases the probability of the miss rate.

### 3) The phase of Artificial Noise (AN) intrusion

The objective of this phase is to reduce the chance of intrusion detection by using the idea of Artificial Noise (AN) to prevent the request of resource allocation to the BS. The AN is transmitted by the intruder to device-2 after the phase of resource spoofing. AN is transmitted instead of the information-bearing signal for the resource allocation. The intruder in this phase acts as a jammer to prevent further communication between device-1 and device-2. The primary aim of incurring AN at device-2 is to impede the further request to BS for not receiving the resource. The AN can be a randomly generated sequence. Considering various demanding competencies of the attack, the proposed attack model shows an immense strength to emerge attack completely. The probability of the successful attempt of the attack is analyzed in the next section.

### C. Impact on miss-rate probability

The miss rate probability of the attack is defined as the parameter to characterize the efficiency and effectiveness of the attack. The approach of attack is based on two possible conditions. The first includes the successful reception of the downlink from the BS by the intruder. The second possibility is when the downlink is not received and is, therefore, missed by the respective intruder. The rate at which the attempts to receive the downlink from the BS are missed by the attacker is known as the miss rate of the attack. The corresponding probability of a miss rate is termed as a miss-rate probability. For the FD attack, the number of missed attempts of an attack is the aggregate of missed attempts in uplink and downlink. In the approach of HD attack, the missed attempts are entirely dependent on the attack of the downlink. Let $E_1^k$ as the total number of attempts made by the intruder to attack the device-1, $m_{dl}^{HD}$ is the number of missed attempts to access resource allocation in HD attack. $m_r^{FD}$ is the number of missed attempts to access resource allocation in the FD (Full-Duplex) attack. The miss rate for the HD attack $M_r^{HD}$ can be expressed as:

$$M_r^{HD} = \frac{m_{dl}^{HD}}{E_1^k} \qquad (21)$$

For Full-duplex the miss rate $M_r^{FD}$ can be obtained as:

$$M_r^{FD} = \frac{m_r^{FD}}{E_1^k} \qquad (22)$$

The above equation can also be re-written as:

$$M_r^{FD} = \frac{m_{ul}^{FD} + m_{dl}^{FD}}{E_1^k} \qquad (23)$$

For the HD attack, the number of missed attempts to receive the downlink successfully is defined for downlink only. Let the number of the missed attempts for the downlink in HD attack be equal to the number of missed attempts for the downlink in FD attack such that equation (23) can be given as:

$$M_r^{FD} = \frac{m_{ul}^{FD} + m_{dl}^{HD}}{E_1^k} \qquad (24)$$

$$= \frac{m_{ul}^{FD}}{E_1^k} + M_r^{HD} \qquad (25)$$

where $m_{ul}^{FD}$ is the number of attempts in the uplink for the FD attack. $m_{dl}^{FD}$ is the number of attempts in downlink for the FD attack, $\frac{m_{ul}^{FD}}{E_1^k}$ is a non-negative quantity whose value lies between 0 and 1. Therefore, from equation (25) it can be concluded as:

$$M_r^{FD} \geq M_r^{HD} \qquad \forall \frac{m_{ul}^{FD}}{E_1^k} \geq 0 \qquad (26)$$

Equation (26) justifies the analysis that the miss rate achieved in the FD attack is more than the miss rate obtained in the HD attack. Therefore, the number of missed attempts to spoof the resources successfully is greater in the FD attack than in the HD attack. Attempting an attack $E_1^k$ times entails a series of independent trials, wherein each one provides two possibilities defined by Bernoulli trials. These two possibilities are termed as a missed attempt and a successful attempt. The probability of the missed attempt in the HD attack is $m_p$. The probability of a successful attempt is $1 - m_p$. Then the probability of the random variable characterized by the miss rate in HD, using binomial distribution is given by:

$$P(M_r^{HD} = m_{dl}^{HD} \ in \ E_1^k \ attempts)$$

$$= \binom{E_1^k}{m_{dl}^{HD}} (m_p)^{m_{dl}^{HD}} (1 - m_p)^{E_1^k - m_{dl}^{HD}} \qquad (27)$$

$$= \frac{E_1^k!}{(E_1^k - m_{dl}^{HD})! \, m_{dl}^{HD}!} (m_p)^{m_{dl}^{HD}} (1 - m_p)^{E_1^k - m_{dl}^{HD}} \qquad (28)$$

The miss rate probability for the full-duplex attack incorporates the miss rate achieved in both uplink and downlink. In general, the probability of the miss-rate for FD attack is given by:

$$P(M_r^{FD} = m_r^{FD} \ in \ E_1^k \ attempts)$$

$$= \binom{E_1^k}{m_r^{FD}} (m_p^{FD})^{m_r^{FD}} (1 - m_p^{FD})^{E_1^k - m_r^{FD}} \qquad (29)$$

Equation (29) can be written as the combination of miss rate probability of downlink and uplink expressed as:

$$P(M_r^{FD} = M_r^{UL} + M_r^{DL})$$

$$= P(M_r^{UL}) + P(M_r^{DL}) - P(M_r^{UL}) \cap P(M_r^{DL}) \qquad (30)$$

The events $P(M_r^{UL})$ and $P(M_r^{DL})$ are independent of each trial and are, therefore, mutually exclusive such that $P(M_r^{UL}) \cap P(M_r^{DL}) = 0$. Therefore, the equation (30) can be given as:

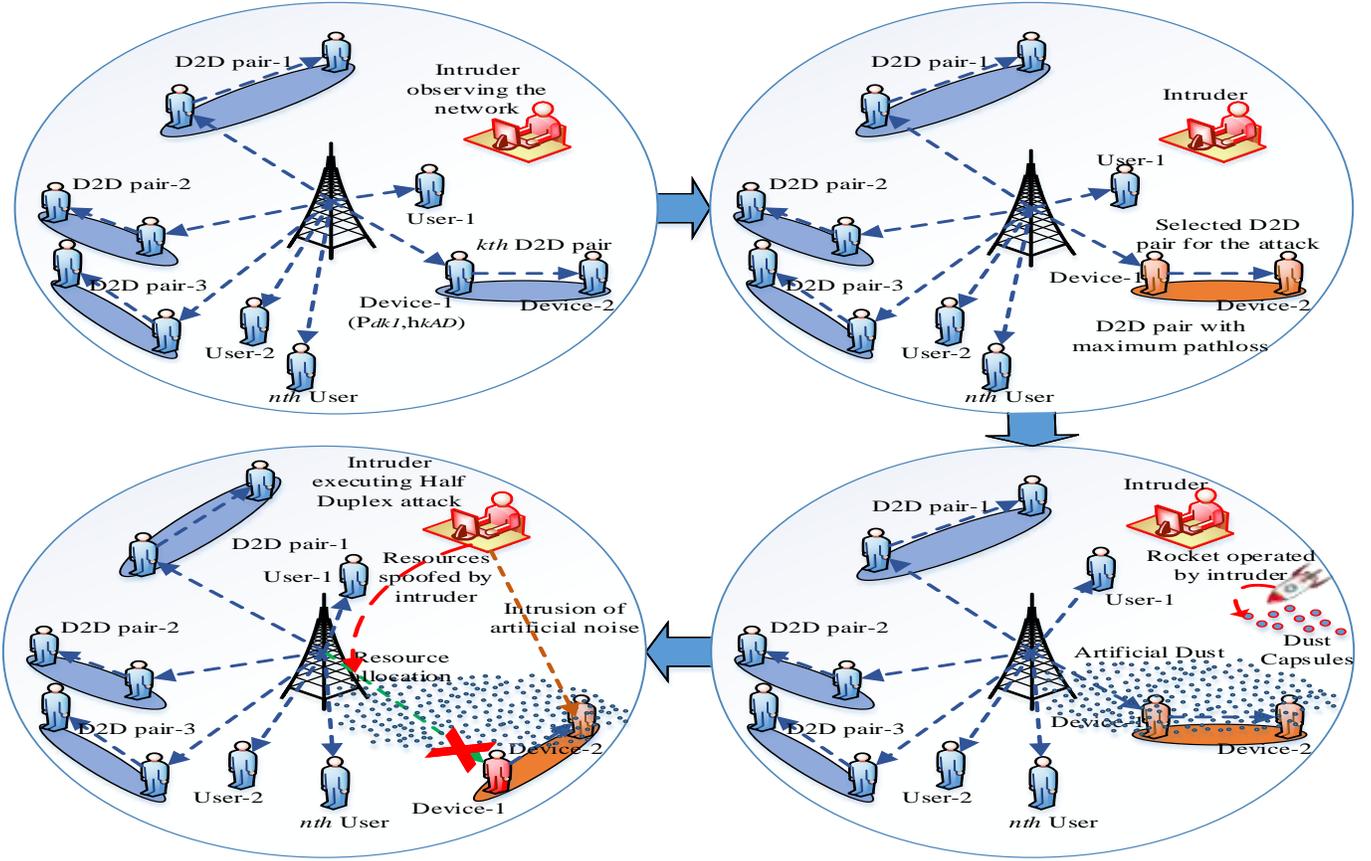

Fig. 2. Process flow of the half duplex attack using an illustrative example

$$P(M_r^{FD}) = P(M_r^{UL}) + P(M_r^{DL}) \qquad (31)$$

Considering equal miss rate probabilities in downlink for HD and FD attack. The HD attack is solely dependent on the downlink. Therefore, $P(M_r^{HD}) = P(M_r^{DL})$. Thus equation (31) can be specified as:

$$P(M_r^{FD}) = P(M_r^{UL}) + P(M_r^{HD}) \qquad (32)$$

For miss-rate probability greater than zero in the uplink for FD attack, the probability cannot be negative. Therefore, equation (32) can also be stated as:

$$P(M_r^{FD}) \geq P(M_r^{HD}) \qquad \forall \, P(M_r^{UL}) \geq 0 \qquad (33)$$

The equation (33) identifies that the miss rate probability of the Full-duplex attack under the impact of AD is always greater than the probability of the miss rate probability of the HD attack. It describes that the chance to spoof the resources successfully from the valid user is more for the HD attack contrary to the FD attack. However, in the case of the none missed attempt of the attack during uplink, the probability of miss rate in FD attack is equal to the probability of miss rate in HD attack.

## IV. Realization and representation of AD (Artificial Dust) based Half-Duplex attack

This section illustrates the algorithm of AD-based HD to provide a complete insight into the approached methodology. The flow charts are shown in appendix (Fig. 13 and Fig. 14.)■ For better understanding, an illustrative example signifying the process flow is represented in Fig. 2. The proposed approach introduces the communication scenario wherein there are $1,2,3, \ldots, k$ D2D pairs, out of which $kth$ D2D pair is initialized for the attack. The attack on the $kth$ D2D pair is chosen based on the degraded channel condition. The $kth$ D2D pair consists of device-1 and device-2. The device-2 is considered to be present farthest from the BS in contrary to device-1, whose signal strength is approachable to the BS. The attacking scenario is defined as an urban or rural communication scenario. The path loss $(p_L^{k1}(n))_{dB}^{AD}$ of the valid user is computed under the attenuation impact of AD. Further, the capacity $C(n)$ is calculated in the observed communication scenario for device-1 of the $kth$ D2D. Similarly, the capacity of the eavesdropper is calculated. Using the estimated capacities of the valid user and the eavesdropper, the secrecy capacity $(C_S(j))_{k1}$ of device-1 from the $kth$ D2D pair is calculated. Where a threshold capacity $C_{th}$ is constrained to achieve secure communication. If the secrecy capacity is below the threshold capacity or the eavesdropper's capacity is more than the valid user then the eavesdropper attempts to attack at the device-1. The given condition achieves the favorable scenario for the attack to occur. As a result, a more convenient attacking approach of HD attack is adapted to spoof the resources. The resources that were meant to be allocated to valid user will be allocated to the attacker. On the other hand, if the condition of favorable scenario does not satisfy such that; the capacity of the valid user is more than the capacity of the intruder or the secrecy capacity of the valid user is greater than the threshold capacity, then the visibility $V$ is decreased till the favorable attacking scenario is achieved. Miss-rate defines the efficiency of the attack based on the number of attempts. For HD attack, the miss rate $M_r^{HD}$ is estimated for the attempts of the attack on the downlink to spoof resources successfully. FD involves the miss rate $M_r^{FD}$ of the attack based on the calculations of attacking attempts involved in uplink and downlink to spoof the resources. On contrary, a comparison is made between the FD attack and the HD attack. Further the probability of miss rate for HD attack $P(M_r^{HD})$ and the FD attack $P(M_r^{FD})$ is calculated using Bernoulli's distribution.

---

**Algorithm 1: Half-Duplex attack under the Artificial Dust (AD) scenario.**

**Input:** Transmission power $P_{Bs}$, Equivalent dust particles radius $r_e$, Frequency $f$, Bandwidth $B$, Visibility $V$ Dielectric constants $\varepsilon'$ and $\varepsilon^*$, $l$ is the user location from BS, $l'$ is the intruder distance, Free space path loss for urban and rural scenario $(p_{D_{k1}})_g$ $g$ =urban, rural scenario, $N \in \mathbb{N}$

**Output:** $(C(n))_{ev}, (C(n))_{K1}, M_r^{HD}, M_r^{HD}(i), P(M_r^{FD}(i))$

1:: $l \leftarrow 0, D_p \leftarrow 1, ev \leftarrow l', define\ g;$

2:: **for** $n = 1$: $N$

3::     compute path loss $(p_L^{k1}(n))_{dB}^{AD}, (p_L)_{ev}$ ;

4::     max $(p_L^{k1}(n))_{dB}^{AD}$ /* select the user with maximum path loss*/;

5:: **end for**

6:: **for** $j = m$: $s$

7::     compute $(\varpi)_{AD}, (\varpi(j))_{ev};$

8::     **for** $n = 1$: $2$

9::         compute capacity ($n = 1 \rightarrow valid\ user, n = 2 \rightarrow eavesdropper$), $(C(n))_j = B\log_2(1 + (\varpi)_n));$

10:: **end for**

11::     $(C_S(j))_{k1} = (C_{k1}(j))_{AD} - (C(j))_{ev};$

12::     **If** $((C_{k1}(j))_{AD} < (C(j))_{ev}$

13::         **or** $(C_S(j))_{k1} < C_{th}$ **then**

14::         $(C(n))_{ev} = B\log_2(1 + (\varpi(n))_{ev})$/* The signal from the BS is received by intruder*/;

15::     **else**

16::         $(C(n))_{K1} = B\log_2(1 + (\varpi(n))_{K1})$/*The signal from the BS is received by device-2*/ ;

17::         Decrement the visibility $V \leftarrow V - -;$

18::     **end if**

19:: **end for**

20:: **for** $i = 1$: $N$

21::     compute HD miss rate $M_r^{HD}, M_r^{HD}(i) = \frac{m_{dl}^{HD}(i)}{N};$

22::     compute probability of $m_{r,HD}(i)$ using Bernoulli's distribution
$$P(M_r^{HD}(i)) = \frac{N!}{\left(N - m_{dl}^{HD}(i)\right)!\, m_{dl}^{HD}(i)!}(m_p)^{m_{dl}^{HD}(i)}\left(1 - m_p\right)^{N - m_{dl}^{HD}(i)};$$

23::     compute miss rate for full-duplex attack
$$M_r^{FD}(i) = \frac{m_{ul}^{FD}(i) + m_{dl}^{FD}(i)}{N};$$

24::     compute the probability of miss rate,
$$P(M_r^{FD}(i)) = P(M_r^{UL}(i)) + P(M_r^{DL}(i)) ;$$

25:: **end for** /* N: maximum value of the parameter*/

26:: **Plot numerical results ;**

---

## V. HALF-DUPLEX ATTACK: AN ILLUSTRATIVE EXAMPLE

The complete procedure of the Half-Duplex (HD) attack under the influence of artificial dust is represented in Fig. 3. A random deployment of 10 users from BS is considered. Based on the distance from the BS, the user with the least channel gain is selected for the evaluation of the attack. The impact of the AD on the propagating signal is obtained from the proposed model on the user farthest from the BS. Depending on the achieved path loss and SNR followed by the estimation of the capacity and secrecy rate is observed. The numeric value at the top of each bar designates the distance of the user from the BS. Fig 3(a) and 3(b) shows the half-duplex and full-duplex attack in the urban scenario under the influence of AD. The methodology is followed for the user with minimum path loss, on which AD impact is observed at the visibility of 0.3Km.

It is analyzed from the observations that user-2 occupies the farthest distance from the BS and possesses the least channel gain. The intruder targets the user-2 to introduce the AD and correspondingly degrade the channel. Also, the deteriorated throughput of user-2 is detected as a consequence of AD attenuation. The incorporated AD attenuation results in the decrease of signal strength of the propagating signal, as shown in Fig. 3(a). The decreased throughput is associated with channel impairments due to the presence of dust particles in the communication channel. For next generation communication networks, the wavelength of the propagating signals are comparable to the dust particle size, which results in the effect of signal attenuation in the form of absorption and scattering, ultimately degrades the capacity of the valid user.

The attack sensitivity under the HD attack and FD attack in an urban communication scenario is depicted in Fig 3(a) and Fig. 3(b). Attack sensitivity defines the measure of attack performance and effectiveness, delineating the minimum capacity of the intruder required to attack successfully. The better the attack sensitivity, the better is the performance of the attack as depicted in Fig. 3. However, from Fig. 3(a) and Fig 3(b), the extent of throughput in the FD attack is superior to the throughput obtained from the HD attack. As the requirement to abandon the secrecy of the valid user necessitates a higher throughput of the intruder than the valid user. From the inspection, it is depicted that the lower value of throughput is executed by the HD attack while higher values of throughput were computed by the attempt of the FD attack. Therefore, it is validated that the sensitivity of the FD attack is less than the sensitivity of the HD attack.

Depending on the achieved capacity by the intruder while executing HD attack a comparison of attack sensitivity in the presence of AD, is made between rural and urban scenarios as shown in Fig. 3(b), Fig 3(d). The acquired capacity of the intruder in a rural scenario possesses more proximity with the capacity of the valid user as shown in $C$ of Fig. 3, to fulfill the

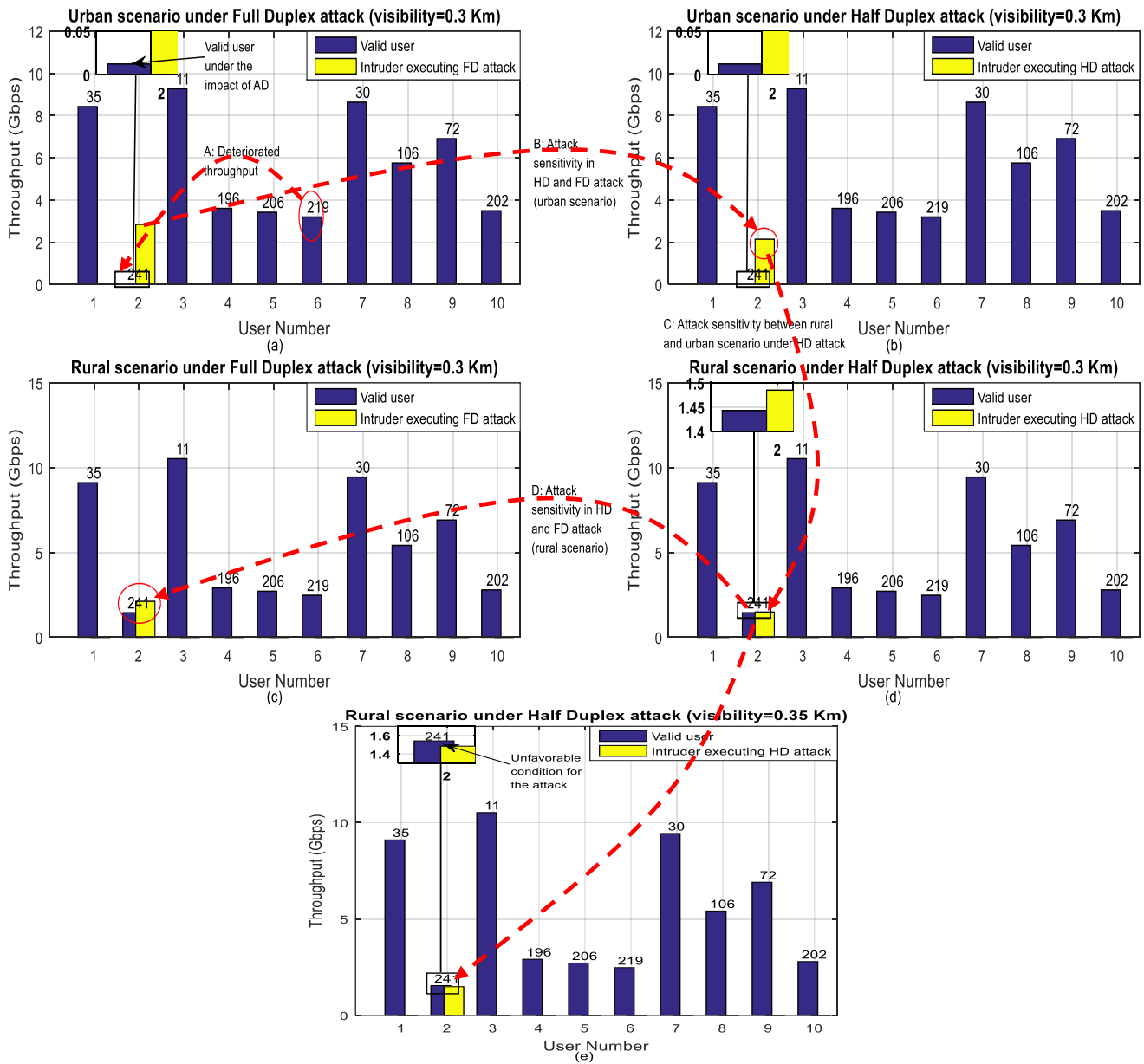

Fig. 3. General analysis of HD (Half-Duplex) attack and FD (Full-Duplex) attack under the impact of AD at the targetted valid user.

.criteria of Wyner's Wiretap theory [19]. Also, in the case of an urban scenario, though the required capacity to formulate an attack on the valid user is the higher capacity of the intruder than the valid user. In our interpretation, it is examined that in an urban scenario the achieved capacity of the intruder is much higher than the required capacity. From the inspection, it is clearly shown that the urban scenario represents less sensitivity contrary to the rural scenario. Therefore, the impact of AD is observed to be dominant in urban scenarios. This is due to the additional propagation losses in the urban scenario, whereas the rural scenario is devoid of such losses. From the observations, it can be concluded that the security attacks based on channel attenuation are less effective in the rural scenario than urban scenario.

The attack sensitivity of the HD attack and FD attack is compared in a rural scenario based on the parameter of the achieved throughput as shown in Fig. 3(c) and Fig. 3(d). It is examined, the achieved capacity of the intruder is just above the capacity of the valid user in both FD and HD attacks. However, in the HD attack, the close proximity of the capacity of the intruder with the required capacity, is observed just above the capacity of the valid user. However, the analysis clearly defines the dominant optimization of the HD attack. In other words, the more adjacent contiguity is defined for HD attack. Moreover, with an increase in sensitivity, the efficiency and effectiveness of the attack are increased. With an increase in efficiency, the source utilization is less for the attacker which ultimately increases the effectiveness of the attack.

The unfavorable scenario of the attack is observed with an increase in visibility. As the increase in visibility decreases the number of particles per unit area, which results in the reduction of propagating attenuation. This is because, more the presence of dust particles, the more is the effect of absorption and scattering. Thus, more will be the attenuation to the propagating signal. From Fig. 3(d) and 3(e), the visibility affects the attacking capability of the intruder. Fig. 3(d) executes the analysis for the operating visibility up to 0.3km. For Fig. 3(e) visibility is 0.35km, which makes the intruder incapable of attempting the attack as the secrecy is maintained due to the increased throughput of the valid user.

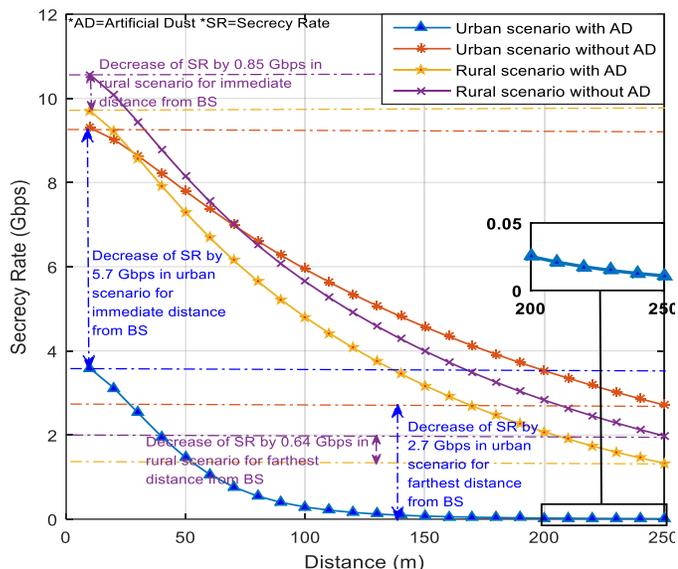

Fig. 4. Secrecy rate versus distance from the BS in the rural and urban scenario

## VI. RESULTS AND DISCUSSIONS

In this section, simulation outcomes based on the HD security attack are examined thoroughly to determine the performance evaluation of a WCN under the impact of AD. The proposed framework of the HD attack involves the spoofing of the resources from the valid user by the intruder. The specified approach ensures more reliability than the FD attack by attempting the solitary selection of downlink. The urban and rural scenarios are taken into consideration for the analysis. The parameters of distance from the BS and the frequency are varied to inspect the characteristics of the secrecy rate. This section is organized into two segments. The first segment defines the simulation background. The second segment outlines the attainment of security parameters in terms of secrecy rate, energy efficiency, complexity, and sensitivity.

### A. Simulation background

The parameters of the simulation are specified in Table IV. The setup of the proposed framework is implemented in MATLAB. The path loss model is defined along with the effect of Rayleigh Flat Fading [17]. The study is observed for the defined parameters of UDN [18]. Graphical analysis is made for the illustrative observation of the achieved results. In the prescribed scenario, an assumption is made that an intruder is capable of achieving the information about the location of the user via real-time satellite tracking, and the users are considered static or moving with a negligible speed for a given period of timestamp.

### B. Results and implications

Two scenarios of rural and urban are independently taken into consideration for the measure of security performance. The parameters of analysis include secrecy rate, distance, operating frequency, sensitivity, energy efficiency, and complexity. The proposed attack analysis is performed under the impact of AD.

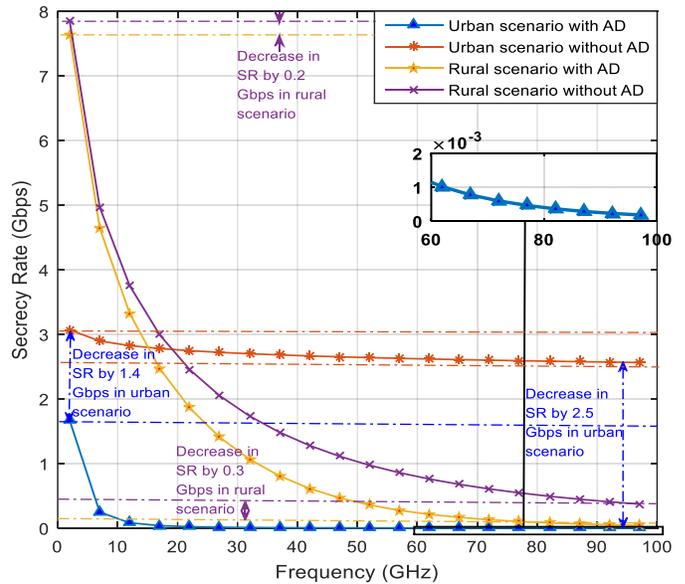

Fig. 5. Secrecy rate versus frequency in the rural and urban scenario

| Parameter | Value | Parameter | Value |
|---|---|---|---|
| Bandwidth (B) | 800MHz | Frequency | 28GHz |
| Noise power | -106dBm | Visibility | 0.3Km |
| Transmission power | 20 mW | Coverage range | 250m |
| Particle radius | 30μm | Distance | 150m |

### 1) Secrecy rate

Fig. 4 shows the relationship between the distance and the secrecy rate. It is observed from the graph that with an increase in distance secrecy rate tends to decrease. However, in the presence of AD secrecy rate tends to decrease drastically. The AD creates an attenuation on the propagating signal resulting in the loss of signal strength in addition to the corresponding path loss as given in equation (5). In the case of rural and urban scenarios, the impact of AD creates a drastic reduction in secrecy capacity confined for urban scenarios contrary to the secrecy capacity achieved in the rural scenario in the presence of AD. For the immediate distance from the BS in the urban scenario, a decreased shift of 5.7Gbps occurs in secrecy rate. For the farthest distance from the BS, the decrease of 2.7Gbps is observed in the presence of AD. Moreover, in the rural scenario, a decrease of 0.85Gbps is observed for the adjacent distance of the valid user from the BS while a reduction of 0.64Gbps is observed for the extreme distance from the BS. Therefore, it can be concluded that the presence of the AD has a substantial impact on the users having a closer distance from the BS, although the path loss is greater for the users having a more considerable distance from the BS without AD. Also, for the urban scenario, the impact of AD is quite considerable. The fact is due to the more prominent propagation loss in the urban scenario.

Fig. 5 illustrates the characteristics of the secrecy rate with an increase in operating frequency, at a distance of 150Km from BS and visibility of 0.3Km. It is observed that with an increase in frequency, the secrecy rate tends to reduce. For rural scenarios, an exponentially decreasing graph is observed with the growing frequency. However, for an urban scenario, a considerably lower decrease in secrecy rate is detected with an increase in frequency. The presence of AD creates a decrease of 1.4Gbps in secrecy rate for urban scenarios and 0.2Gbps for rural scenarios at an operating frequency of 2.1GHz.

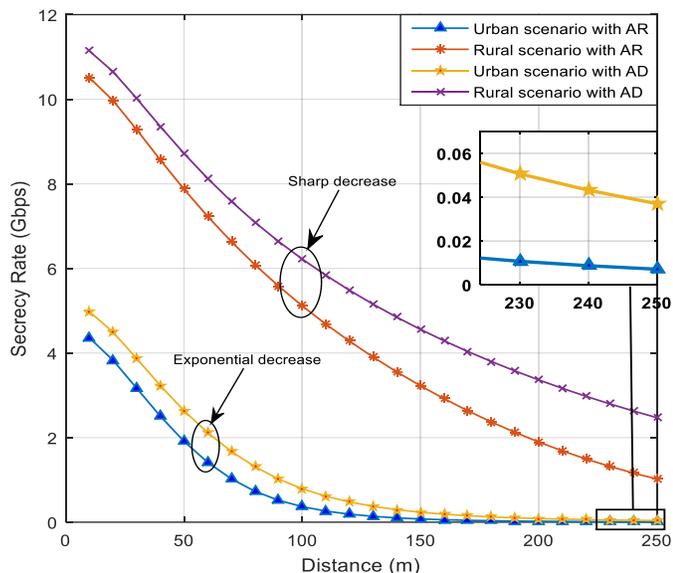

Fig. 6. Secrecy Rate versus distance in rural and urban scenario in the presence of AR (Artificial Rain) and AD (Artificial Dust)

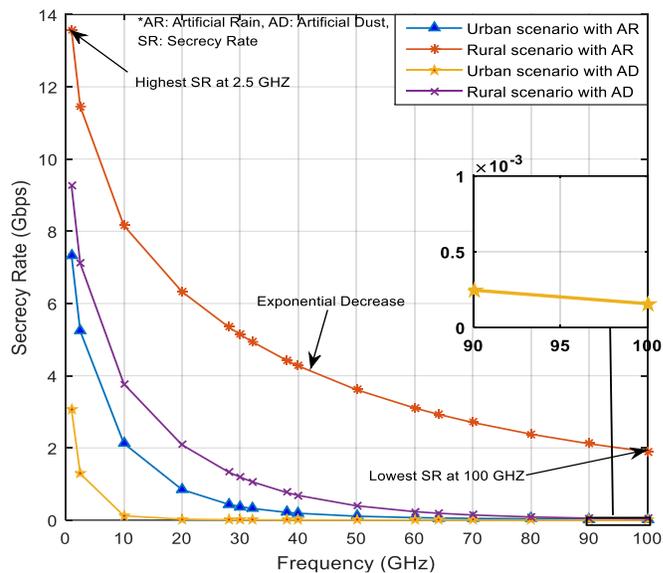

Fig. 7. Secrecy rate versus frequency for an urban and rural scenario in the presence of AR (Artificial Rain) and AD (Artificial Dust)

However, at the higher millimeter-wave frequency of 100GHz, a shift of 2.5Gbps reduction in secrecy rate for urban scenario and 0.3Gbps for rural scenario is encountered. Hence, the execution of AD has a more significant impact at higher frequencies specifically at millimeter waves for the next generation WCN. In view of the urban scenario, the attenuation is comparatively more than in the rural scenario. One of the reasons is the existence of dense obstacles present in the urban communication scenario. The penetration loss in millimeter waves is observed to be more which leads to the fall of secrecy rate.

Fig. 6. manifests the variation of secrecy rate with respect to the distance under the influence of AR [20] and AD. It is quite clear from the analysis that with an increase in distance, the secrecy capacity tends to decrease. This effect is due to the decrease in signal strength with an increasing distance. The effect of AD is observed to be less significant than the effect of AR. The statistics are such because of the more penetration loss in AR depending on the factors of particle size and the concentration of the AR droplets in contrary to the AD particles. Approaching the characteristics of the rural and urban communication scenario, an exponentially gradual decrease of secrecy rate concerning the increasing distance is observed while the sharp decrease of secrecy rate is detected in the rural scenario.

For the rural scenario, the impact of AR (Artificial rain) is dominant on the secrecy rate as compared to the AD for varying distances. Also, in the urban communication network, the effect of the AR is greater than the impact of AD in the communication network with varying distances of the valid user from the BS.

Fig. 7. evinces the comparative distinctive analysis of secrecy rate concerning an increasing frequency in the presence of an AR and AD. Exponentially decreasing specifics of secrecy rate are witnessed in corresponding to the rising frequencies. The presence of AR and AD creates the attenuation in the transmitted signal in a communication network, which ultimately creates an additional reduction in the secrecy rate. The size of particles is comparable to the wavelength, which gives rise to the scattering such that the wavelength is inversely proportional to the scattering intensity. It correspondingly gives an ultimate effect on the frequency due to the inverse relation of frequency and wavelength. Also, it provides the direct relationship between frequency and the intensity of the scattering. Therefore, the impact of AD attenuation increases with an increase in frequency. It indicates that the communication networks operating at higher frequencies or the mmWave communication signals are more affected by the obstruction than low-frequency signals.

*Remarks 2:* *The next generation communication network of 5G NR (New Radio)/6G is more prone to AD attenuation. Therefore, more vulnerable to the HD attack as an impact of higher frequencies.*

One of the significant consequences of the smaller wavelength is the higher penetration loss during the transmission via certain materials. The effect of AD is detected to be preeminent in comparison to AR present in the communication network for both urban and rural scenarios. It is evident from the analysis, the rural communication scenario under the influence of AR possesses the highest secrecy rate than the secrecy rate achieved in urban scenarios in the presence of AR. Also, the secrecy rate under AD is lower than the AR. Thus, for mmWave frequencies, the security of the urban scenario is compromised in the presence of AD and is, therefore, more prone to security attacks.

Fig. 8. outlines the comparative attacking ability of the intruder for the range of frequencies. The user is considered under the influence of AD at a distance of 150Km from the BS. It is observed from the graph that the intruder at lower frequencies occupies inadequate capacity to impose an attack. In other words, the capacity of the intruder is less than the capacity of the valid user at lower frequencies. However, at higher frequencies in the range of mmWaves, the impact of the AD attenuation becomes dominant. Thus, the capacity of the

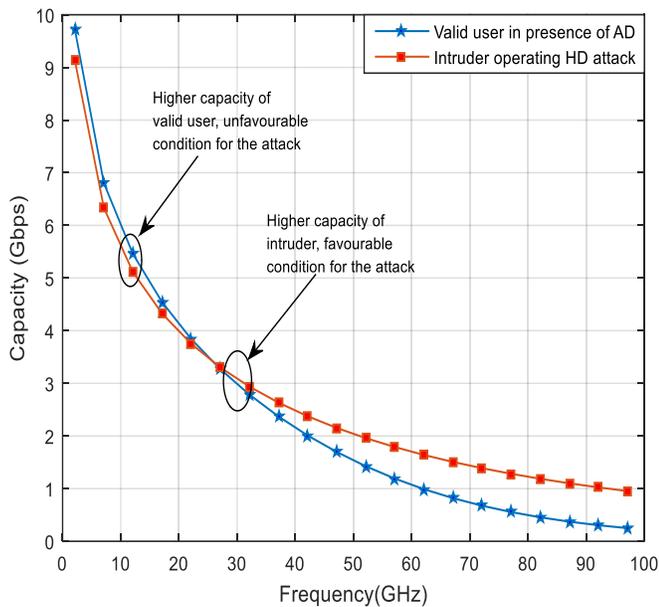

Fig. 8. Capacity versus frequency for the intruder and valid user under HD attack

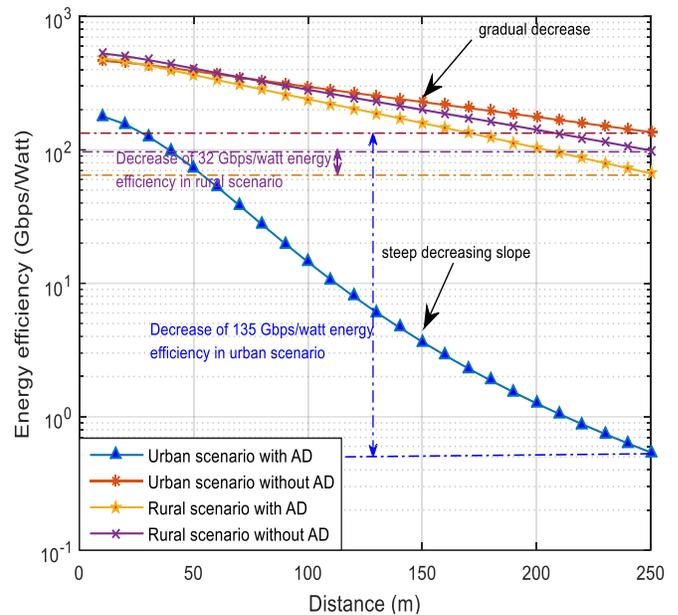

Fig. 9. Energy efficiency versus distance for the rural and urban scenario in the presence of AD

valid user decreases below the capacity of the intruder. Hence, at higher frequencies of next generation WCN, the intruder while incorporating AD on the valid user is capable of eavesdropping.

***Theorem 1:*** *The operating frequency has a direct impact on the attenuation due to AD in the form of absorption and scattering*

$$f \propto \phi_{abs} \tag{34}$$

$$f \propto \phi_{sc} \tag{35}$$

***Proof:*** The frequency can be expressed as:

$$f = \frac{c}{\lambda} \tag{36}$$

or

$$f \propto \frac{1}{\lambda} \tag{37}$$

$$f^4 \propto \frac{1}{\lambda^4} \tag{38}$$

Also, scattering efficiency $\phi_{sc}$ [15] can be given as:

$$\phi_{sc} \propto \frac{1}{\lambda^4} \tag{39}$$

And absorption efficiency $\phi_{abs}$ [16] is given by:

$$\phi_{abs} \propto \frac{1}{\lambda} \tag{40}$$

From equation (38) and (39)

$$f^4 \propto \phi_{sc} \tag{41}$$

From equation (37) and (40)

$$f \propto \phi_{abs} \tag{42}$$

From the analysis, it is observed that frequency is directly proportional to the scattering and absorption. Therefore, an increase in frequency, attenuation increases due to an increase in scattering and absorption efficiencies of the particles present in the communicating channel. ∎

### 2) Energy Efficiency

Fig. 9. demonstrates the traits of energy efficiency corresponding to the distance of the user from the BS. The energy efficiency shows a declining behavior with respect to the distance of the user from the BS. It derives from the fact that with an increase in distance the power consumption tends to increase for the transmission of signals to the destination. For urban scenarios, energy efficiency drops significantly by a factor of 135Gbps/Watt in the presence of AD in contrary to the rural scenario where the reduction takes place by a factor of 32Gbps/Watt. Therefore, it is revealed that a gradual decrease occurs in energy efficiency for higher frequencies greater than 27GHz under rural scenarios while an abrupt decrease of energy efficiency is identified in urban scenarios. However, with an increment in the distance, the difference between the energy efficiency obtained in the presence and the absence of AD tends to upturn extensively for an urban scenario contrary to the rural scenario.

### 3) Sensitivity, miss rate probability, and complexity

The procedure of attack sensitivity is illustrated to monitor the efficiency of the attack. It governs the minimum optimal value of capacity required by the intruder to trigger the attack. It is based on Wyner's Wiretap theory [19], where the security of the information is preserved if the channel between the valid user and BS is of higher quality than the channel between BS and eavesdropper. Based on this approach, the execution of an attack can be operated by optimizing the channel capacity of intruders just greater than the channel capacity of the valid user. However, the parameter of sensitivity in our approach defines the capability of the intruder to optimize the minimal channel capacity just above the capacity of the valid user. The formula for the attack sensitivity is derived from the capacity of the user under attack and the capacity of the attacker. The derived equation for the sensitivity of an attack is interpreted as a closed-form expression. $d'$ denotes the capacity of the user under the influence of AD. $d$ denotes the difference between the capacity of the user and intruder under the AD effect. The sensitivity of the attack can be expressed as:

$$\alpha = \frac{d' - d}{d'} \tag{43}$$

Sensitivity is a unitless quantity and computes the performance analysis of the attack. Sensitivity has a high impact on the efficient usage of intruder resources such as power. The more the sensitivity of the intruder, the more optimal value of intruder capacity close to the capacity of the valid user is detected, the less would be the intruder resource usage. Consequently, optimal sensitivity reduces the requirement of capacity and power of the intruder. Therefore, the mechanism detects more efficiency of the attack for more sensitivity (**see lemma 1**). ∎

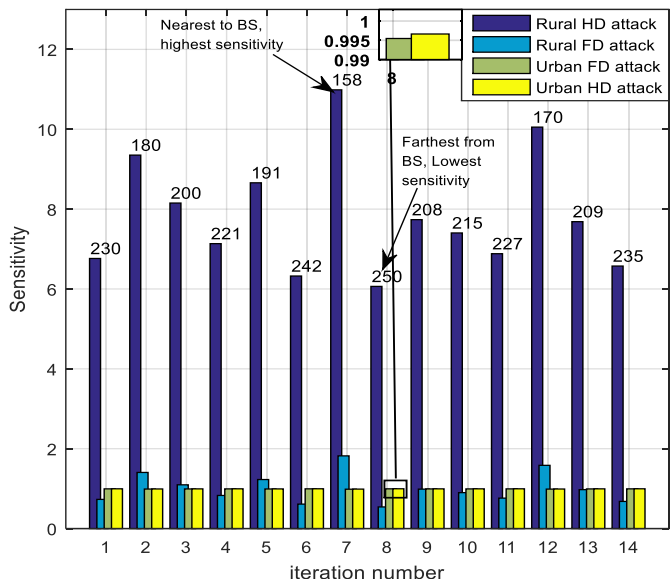

Fig. 10. Sensitivity analysis of HD and FD attack in the rural and urban scenario under the impact of AD

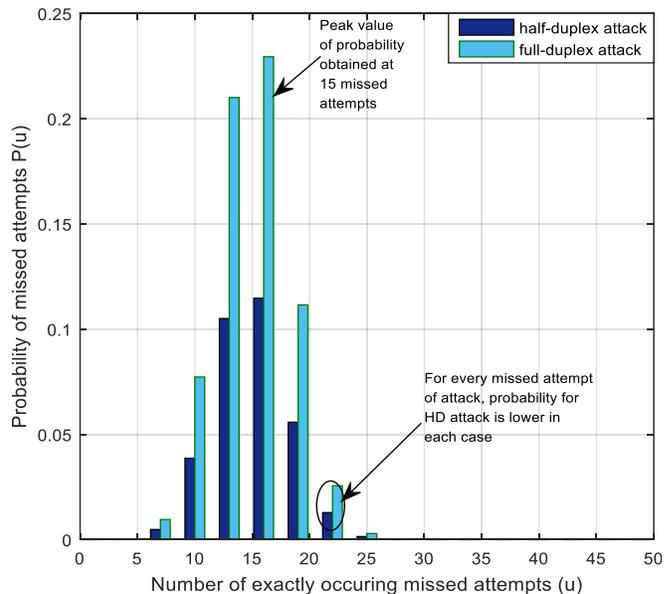

Fig. 11. Probability analysis of missed attempts for HD and FD

Fig. 10. states the sensitivity illustration for the users under attack for 14 iterations. The distance of the user from the BS on which the intruder incorporates the attack is shown on the top of each bar. For each iteration, the user with the most degraded CSI is targeted for the attack. The valid user is incorporated under the influence of AD at visibility of 0.3Km. The inquiry is evaluated for urban and rural scenarios incorporating HD and FD attacks. It is observed from the graph that the sensitivity is more effective in the rural scenario than in the urban scenario. Correlating the attacking mechanisms of HD and FD attack, the HD attack exhibits dominant sensitivity than the sensitivity achieved from the FD attack. Sensitivity is observed to be interdependent on the distance defined as the users closer to the BS obtains immense sensitivity, and the users farthest from the BS attain the least sensitivity correspondingly. Therefore, it can be concluded that sensitivity is inversely proportional to the distance from the BS. This is because the signal strength decreases with an increase in distance from the BS.

Fig.11. shows the probability analysis of the HD and FD attacks for the missed attempts of the attack. The probability of the hit attempt of an attack is taken as 0.3 while the probability of the missed attempt of an attack is taken as 0.7 for the analysis. The characteristics of the discrete distribution are observed to be similar to a bell shape curve. The occurrence of 15 missed attempts from 50 attempts of an attack is observed to attain the probability of 0.227, which is the maximum probability. It is depicted from the graph, that with an increase in the number of the occurrence of the attempts, the probability correspondingly tends to decrease. Furthermore, for the equal probability of missed attempt and hit attempt viz 0.5, a similar characteristic of the graph is observed with a shift from the axis of the number of occurrences of missed attempts. Also, from Fig. 11, it is reasonably apparent that the probability of the

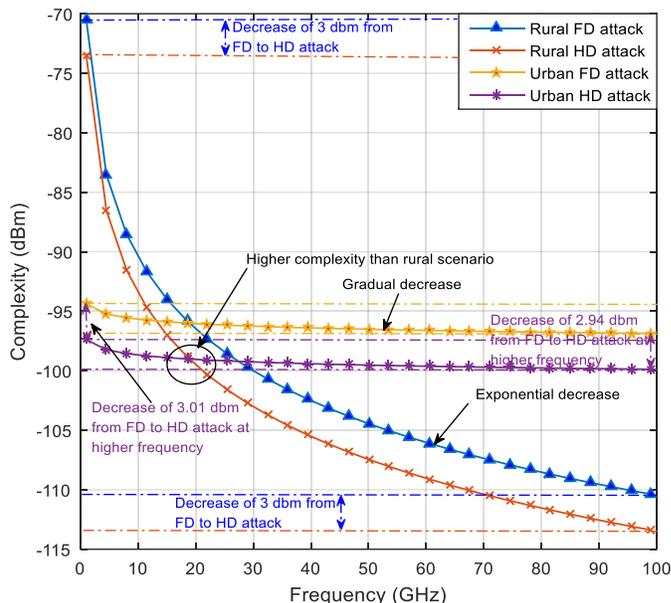

Fig. 12. Complexity versus frequency for HD (Half-Duplex) attack ad FD (Full Duplex) attack under rural and urban scenario

missed attempts is greater in FD attacks than in the HD attack as the HD attack incorporates downlink only, instead of targeting both the downlink and uplink as in FD attack.

Fig. 12. represents the validation of the proposed attack based on the complexity parameter with an increasing frequency. The proposed scheme emphasizes AD to decrease the secrecy rate of the communication network. It is depicted that the complexity of the attack decreases with an increase in frequency. It is due to the fact that with an increase in frequency the secrecy rate tends to decrease therefore, less energy or power is required to establish an attack at higher frequencies Thus, complexity provides an inverse impact on reliability and efficiency. The more complexity, the less is the reliability. For HD attack, the attained complexity is lower than the complexity of the FD attack for the increasing frequency. Contrariwise, in the rural scenario, the FD attack provides higher complexity than the complexity provided by the HD attack. Similarly, the HD attack in an urban communication network occupies less complexity than the FD attack for increasing frequencies including mmWave frequencies. However, at frequencies greater than 20GHz the complexity of the urban scenario is observed to be more than the rural scenario. Moreover, the urban scenario under HD attack for millimeter-wave frequencies shows less complexity than urban scenarios under FD attack. From the analysis, it can be concluded that, for the frequencies higher than 20GHz, the urban scenario is more inclined to security attacks in the presence of AD.

## VII. Conclusion and Future Work

The security impact of the AD (Artificial Dust) is addressed in this paper, followed by the execution of the HD (Half-Duplex) attack. The estimation of the AD attenuation is obtained from the proposed dust model. Based on the calculated attenuation, the secrecy capacity of the valid user is evaluated. Due to the reduced secrecy capacity intruder attempts the HD (Half-Duplex) attack. The performance of the HD attack is then compared with the FD attack. The attacks were executed in rural and urban scenarios. The evaluation of the attack was executed based on miss-rate. Miss-rate defines the effectiveness of the attack. Miss-rate is expressed in terms of Bernoulli's distributions and Poisson's process. Moreover, the parameter of attack sensitivity is defined to determine the efficacy of the attack. Finally, simulation results were achieved to illustrate the proposed scheme.

Furthermore, the efficient resource utilization of the attacker can be improvised by the implementation of training models. The proposed scheme can be extended to the broad area of UAV applications for the remote area defense applications to achieve the intended information from the opponent. Additionally, for more accuracy of the proposed scheme, the real-time based data and the estimated data are required to be analyzed. However, apart from the numerical differences, the characteristic trends for both the real data and estimated data will be similar.

## Appendix

■ *Lemma 1:* The attack sensitivity of an intruder is defined by the minimum capacity employed by the intruder to attempt a successful attack. The performance of the attack in terms of attack sensitivity is given by:

$$S_{ev} = \frac{d' - d}{d'} \qquad (44)$$

*Proof:* Generally, the performance of the communication system is defined by the parameter of receiver sensitivity. It is the minimum received signal power at a specific Bit Error Rate (BER) to designate the quality of the service at the receiver given by:

$$S_r = I_{MDS} + \varpi \qquad (45)$$

where $I_{MDS}$ is the Minimum detectable Signal Power and can be expressed as:

$$I_{MDS} = -174_{dBm} + 10\log_{10} B + N_f \qquad (46)$$

Or

$$S_r = -174_{dBm} + 10\log_{10} B + N_f + \varpi \qquad (47)$$

where $N_f$ is the noise figure of the receiver, $\varpi$ is the SNR of the receiver, $B$ is the bandwidth of the receiver. Considering the attack sensitivity of the intruder, based on the concept of Wyner's theory of wiretap channel, the parameter of capacity is

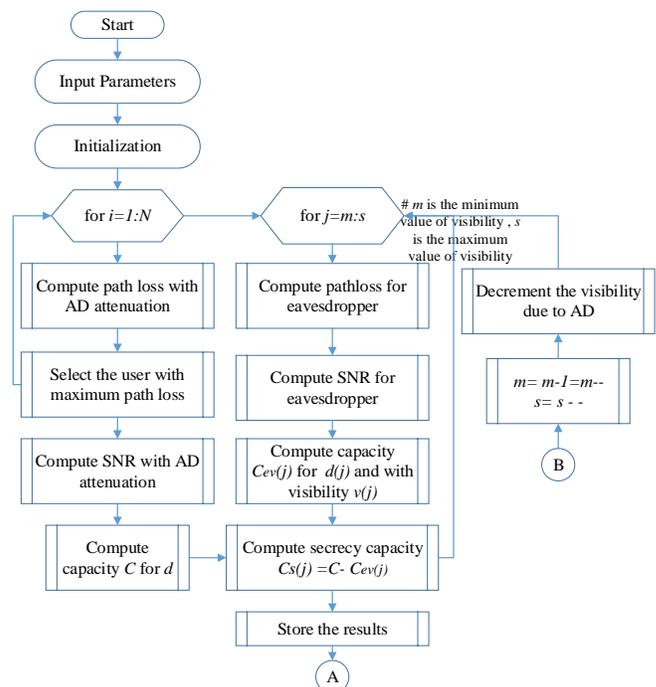

Fig. 13. Flow chart for secrecy capacity under the impact of AD

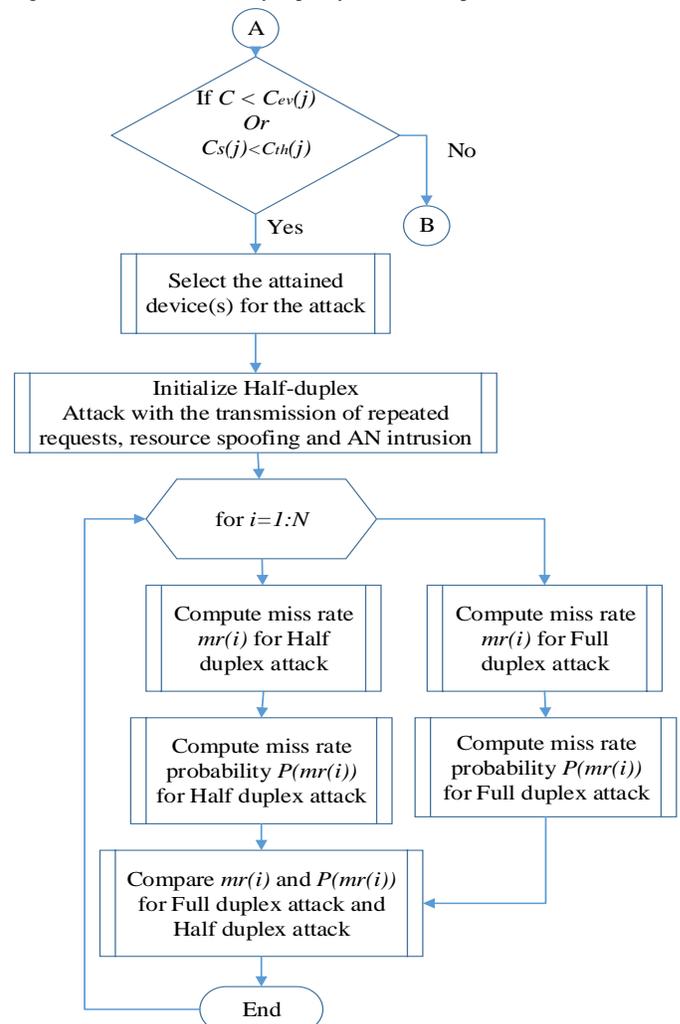

Fig. 14. Flow chart for the Half-duplex attack

taken into consideration such that the attack sensitivity of an intruder is signified as:

$$S_{ev} = \frac{d_o - d'}{d'} \tag{48}$$

where,

$d_o = B \log_2(1 + \varpi_{d_o}), d' = d - d_o$, and $d = B \log_2(1 + \varpi_d)$

$$S_{ev} = \frac{2d_o - d}{d - d_o} \tag{49}$$

Or

$$S_{ev} = \frac{2B \log_2(1 + \varpi_{d_o}) - B \log_2(1 + \varpi_d)}{B \log_2(1 + \varpi_d) - B \log_2(1 + \varpi_{d_o})} \tag{50}$$

Or

$$S_{ev} = \frac{2 \log_2(1 + \varpi_{d_o}) - \log_2(1 + \varpi_d)}{\log_2(1 + \varpi_d) - \log_2(1 + \varpi_{d_o})} \tag{51}$$

$$S_{ev} = \frac{-\left\{ \log_2\left[ \frac{1 + \varpi_d}{(1 + \varpi_{d_o})^2} \right] \right\}}{\log_2 \left[ \frac{1 + \varpi_d}{1 + \varpi_{d_o}} \right]} \tag{52}$$

Using the logarithmic base change rule:

$$S_{ev} = -\left\{ \log_{\left( \frac{1 + \varpi_d}{1 + \varpi_{d_o}} \right)} \left[ \frac{1 + \varpi_d}{(1 + \varpi_{d_o})^2} \right] \right\} \tag{53}$$

Or

$$S_{ev} = -\log_{\left( \frac{1 + \varpi_d}{1 + \varpi_{d_o}} \right)} [1 + \varpi_d] + 2 \log_{\left( \frac{1 + \varpi_d}{1 + \varpi_{d_o}} \right)} [1 + \varpi_{d_o}] \tag{54}$$

$$S_{ev} = \varpi'_d + \varpi'_{d_o} \tag{55}$$

where $\varpi'_d = -\log_{\left( \frac{1 + \varpi_d}{1 + \varpi_{d_o}} \right)} [1 + \varpi_d]$ defining the minimum eavesdropping capacity and $\varpi'_{d_o} = 2 \log_{\left( \frac{1 + \varpi_d}{1 + \varpi_{d_o}} \right)} [1 + \varpi_{d_o}]$ is the function of the SNR. Therefore, finding the analogy of equation (55) with the equation (45) of receiver sensitivity verifies the equation of attack sensitivity in equation (48)

■ The flow chart of the proposed mechanism is shown in Fig. 13 and Fig. 14. Secrecy rate estimation is represented in Fig. 13. The attack analysis is shown in Fig. 14.

## Acknowledgment

The authors gratefully acknowledge the support provided by 5G and IoT Lab, DoECE, and TBIC, Shri Mata Vaishno Devi University, Katra, Jammu and Kashmir, India.

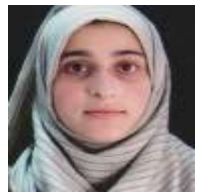

**Misbah Shafi** has received the B.E degree and M.Tech degree in Electronics and Communication Engineering in 2016 and 2018. She is currently pursuing a Ph.D. degree in Electronics and Communication Engineering at SMVD University, J&K. Her research interest includes network security and wireless communication. Currently, she is doing her research on security issues of 5G, 5G NR, and 6G.

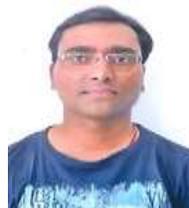

**Dr. Rakesh K Jha (S'10, M'13, SM 2015)** is currently an associate professor in the department of Electronics and Communication Engineering, Indian Institute of Information Technology Design and Manufacturing Jabalpur, India. He has also worked as an associate professor at SMVD University, J&K, India. He is among the top 2% researchers of the world. He has published more than 41 Science Citation Index Journals Papers including many IEEE Transactions, IEEE Journal, and more than 25 International Conference papers. His area of interest is Wireless communication, Optical Fiber Communication, Computer Networks, and Security issues. Dr. Jha's one concept related to the router of Wireless Communication was accepted by ITU in 2010. He has received the young scientist author award by ITU in Dec 2010. He has received an APAN fellowship in 2011, 2012, 2017, and 2018 and a student travel grant from COMSNET 2012. He is a senior member of IEEE, GISFI and SIAM, International Association of Engineers (IAENG) and ACCS (Advance Computing and Communication Society). He is also a member of, ACM and CSI, with many patents and more than 1941 citations on his credit.